\documentstyle[12pt]{article}
\def\phii{\mbox{$\phi$}}

\def\n{\mbox{$N$}}
\def\ua{\mbox{$U(1)$}}
\def\upq{\mbox{$U(1)_{PQ}$}}

\def\mm{\mbox{$\overline{M}$}}
\def\npb#1#2#3{\mbox{Nucl. Phys. {\bf B#1} (#2) #3}}
\def\plb#1#2#3{\mbox{Phys. Lett. {\bf B#1} (#2) #3}}
\def\prd#1#2#3{\mbox{Phys. Rev. {\bf D#1} (#2) #3}}
\def\prl#1#2#3{\mbox{Phys. Rev. Lett. {\bf #1} (#2) #3}}
\def\prc#1#2#3{\mbox{Phys. Rep. {\bf #1} (#2) #3}}

\def\ijmp#1#2#3{\mbox{Int. J. Mod. Phys. {\bf A#1} (#2) #3}}

\begin{document}

\baselineskip 24pt

\newcommand{\sheptitle}
{A Next-to-Minimal Supersymmetric Model \\ of Hybrid Inflation}

\newcommand{\shepauthor}
{M. Bastero-Gil$^\ast $ and S. F. King$^{\dagger}$
\footnote{On leave of absence from $^\ast$.} }

\newcommand{\shepaddress}
{$^\ast$Department of Physics and Astronomy,
University of Southampton, \\ Southampton, SO17 1BJ, U.K.\\
$^\dagger$ Theory Division, CERN, CH-1211 Geneva 23, Switzerland}

\newcommand{\shepabstract}
{We propose a model of inflation based on a simple variant of the NMSSM,
called $\phi$NMSSM, where the additional singlet $\phi$
plays the role of the inflaton in hybrid (or inverted hybrid) type models. 
As in the original NMSSM, the $\phi$NMSSM solves the
$\mu$ problem of the MSSM via the VEV of a gauge singlet $N$,
but unlike the NMSSM does not suffer from domain wall problems
since the offending $Z_3$ symmetry is replaced by an approximate
Peccei-Quinn symmetry which also solves the
strong CP problem, and leads to an invisible axion with 
interesting cosmological consequences.
The PQ symmetry may arise from a superstring model 
with an exact discrete $Z_3 \times Z_5$ symmetry after compactification.
The model predicts a spectral index $n=1$ to one part in $10^{12}$.}

\begin{titlepage}
\begin{flushright}
CERN-TH/97-262\\
hep-ph/9709502\\
\end{flushright}
\vspace{.1in}
\begin{center}
{\large{\bf \sheptitle}}
\bigskip \\ \shepauthor \\ \mbox{} \\ {\it \shepaddress} \\ \vspace{.5in}
{\bf Abstract} \bigskip \end{center} \setcounter{page}{0}
\shepabstract
\begin{flushleft}
CERN-TH/97-262\\
\today
\end{flushleft}
\end{titlepage}

There is to date no standard model of inflation, and although there has
been a good deal of progress in recent years in this area much of 
the current activity 
has been concerned with conceptualised field theoretic
models rather than well motivated particle physics based models
\cite{kolbturner}.
Possibly the best motivated particle physics model beyond the
standard model is the minimal supersymmetric standard model (MSSM).
However the only Higgs fields in the MSSM are the two doublets
$H_1,H_2$, which develop vacuum expectation values (VEVs) of order the
weak scale, and it is very difficult if not impossible to develop
a model of inflation using only these fields for several reasons.
The primary reasons are that the electroweak scale turns out to
be too small and the Higgs potential is not sufficiently flat.
The so called next-to-minimal supersymmetric standard model
(NMSSM) is more promising from the point of view of inflation
since it contains, in addition to the two Higgs doublets, a Higgs 
singlet $N$ which may develop a large VEV. 

The usual NMSSM does not require the 
$\mu H_1H_2$ term of the MSSM, replacing it with
a $\lambda NH_1H_2$ term, and thereby solving the
$\mu$ problem\footnote{Note that the Giudice-Masiero mechanism \cite{mu}
presents a solution to the
$\mu$ problem within the MSSM
by generating the $\mu$ term via a non-minimal
Kahler potential.}.  
The NMSSM also involves a
term $k N^3$ in the superpotential
so that the model has an exact $Z_3$ symmetry
\cite{NMSSM,NMSSMphenom}. However this is broken at the weak scale
leading to a serious domain wall problem \cite{walls,NMSSMwalls}.
Originally it was thought
that the $Z_3$ may be slightly violated by Planck scale operators,
leading to a pressure term that removes the walls. However without an exact
$Z_3$ symmetry supergravity tadpole diagrams will lead to a large
singlet mass in the low energy theory, and the amount of $Z_3$ breaking
required to solve the domain wall problem is in conflict with
requirement that tadpoles do not make the singlet too heavy 
\cite{NMSSMwalls,various}. 

It transpires that, without fine-tuning, the NMSSM does not lead
to a sufficiently flat potential along which the inflaton may roll.
In order to overcome this we introduce a second singlet $\phi$,
and replace the term $N^3$ in the NMSSM by $\phi N^2$.
Thus our model is based on the superpotential:
\begin{equation}
W_{\phi NMSSM} =\lambda NH_1H_2 -k \phi N^2
\label{phiNMSSM}
\end{equation}
Note that our model has the same number of dimensionless couplings as the
original NMSSM, and we have used the same notation $\lambda ,k$ to emphasise 
this. With this modification the
field $\phi$ appears only linearly in the
superpotential and so will have a very flat potential, lifted only by
a tiny mass $m_{\phi}$ of order electronvolts, and will play the role
of the inflaton field of hybrid inflation 
\cite{hybrid,copeland,morehybrid} if 
$m_{\phi}^2>0$ or inverted hybrid inflation \cite{inverted} if
$m_{\phi}^2<0$. In the case of inverted hybrid inflation the present
model provides an interesting counter example to the problems
raised in Ref. \cite{invertedproblems}.
Inflation ends when $\phi$ reaches a critical value
$\phi_c \sim 10^{13}$ GeV after which the $N$ field, which has a zero
value during inflation, develops a VEV $<N> \sim \phi_c$.
Interestingly the inflaton also develops an eventual VEV
$<\phi > \sim \phi_c$ via a tadpole coupling, which is 
typical of inverted hybrid inflation but quite extraordinary for
hybrid inflation.The resulting dimensionless couplings are
$\lambda , k \sim 10^{-10}$, whose smallness will be explained
by embedding the model into a string inspired model where the couplings 
result from higher dimension operators, controlled by discrete symmetries.
Note that radiative corrections to the inflaton mass are controlled
by $\lambda , k$ and are of order the inflaton mass itself.

Having replaced the NMSSM superpotential by 
Eq. (\ref{phiNMSSM}), the troublesome $Z_3$ symmetry 
is replaced by a global $U(1)_{PQ}$ Peccei-Quinn symmetry
where the global charges of the fields satisfy:
\begin{equation}
Q_N+Q_{H_1}+Q_{H_2}=0, \ \ Q_{\phi}+2Q_N=0.
\label{PQcharges}
\end{equation}
with the quark fields having the usual axial PQ charges.
The global symmetry forbids additional couplings such as $N^3$, $\phi H_1H_2$
and so on, but is broken at the scale of the VEVs releasing a very light
axion. The axion scale $f_a$ is therefore of order $\phi_c$ in this model.
The axion will be an invisible 
Dine-Fischler-Srednicki-Zhitnitskii (DFSZ) \cite{dfsz} type axion,
which couples to ordinary matter through its mixing with the standard
Higgses after the electroweak phase transition.
Once we embed our model into a string motivated model, the global
PQ symmetry will emerge as an approximate accidental symmetry of an underlying
discrete symmetry, and we need to discuss such questions as the solution
to the strong CP problem in this wider context.
Note that if we had simply 
removed the $N^3$ term from the NMSSM superpotential
and not replaced it with anything
then the theory would also have a PQ symmetry,
and the potential would also be flat in the $N$ direction, 
and then one might be tempted to identify $N$ with the
inflaton of hybrid inflation. However in such a scenario the height of the
potential during inflation would be of order 1 TeV, leading to
an inflaton mass very much smaller than the radiative corrections
to its mass of order 1 eV, which would require unnatural fine-tuning. 
By contrast, with the $\phi N^2$ term present,
the height of the potential during inflation is about $10^8$ GeV
and the COBE constraint may be satisfied by an inflaton mass of about 1 eV
which is the same order as the radiative corrections to its mass,
leading to a natural scenario with no fine-tuning required.

The tree-level potential which follows from the superpotential in 
Eq. (\ref{phiNMSSM}) can be written, 
if we ignore $H_1,H_2$ which have smaller VEVs,
\begin{eqnarray}
V_0 & = & V(0)+V(\phi ,N) \nonumber \\
  V(\phi,N)&=& k^2 N^4 + m^2(\phi) N^2 +m_{\phi}^2\phi^2
\,, \label{vphin2}
\end{eqnarray}
with the field dependent $N$ mass given by,
\begin{equation}
  m^2(\phi)= m_N^2- 2 k A_k \phi + 4 k^2 \phi^2 \,.
\end{equation}
We have taken \phii\ and \n\ to be the real components of the
complex singlets, and included the soft breaking parameters from the
soft supersymmetry breaking potential terms
$m_NN^2$, $m_{\phi}\phi^2$ and $A_kk\phi N^2$.
We have also added by hand a constant vacuum energy $V(0)$ to the potential,
about which we shall say more later.
Note that $m^2(\phi)=0$ for
\phii\ equal to a critical value\footnote{We require that the condition
$A_k^2 > 4 m_N^2$ is fulfilled.}: 
\begin{equation}
 \phi_c^{\pm}=\frac{A_k}{4 k} \left( 1 \pm \sqrt{1-4 \frac{m_N^2}{A_k^2}}
 \right) \,.
\label{phic}
\end{equation}

In order to discuss inflation we need to specify the sign of
the inflaton mass squared $m_{\phi}^2$. If $m_{\phi}^2>0$ 
(as in hybrid inflation) then,
for $\phi > \phi_c^+$, \n\ will be driven to a local
minimum (false vacuum) with \n=0. 
Having a positive mass squared, $\phi$ will roll towards the origin
and $m^2(\phi)$ will become
negative once the field \phii\ reaches $\phi_c^+$.
After that, the potential develops an instability in the \n=0
direction, and both singlets roll down towards the global minimum, 
\begin{eqnarray}
   <\phi>&=& \frac{A_k}{4 k} \,,\\
    <N>  &=& \frac{A_k}{2\sqrt{2} k}\sqrt{1-4 \frac{m_N^2}{A_k^2}}
=\sqrt{2} \left|
   \phi_c^{\pm}-<\phi > \right|\,,
\end{eqnarray}
signaling the end of the inflation. 
On the other hand if $m_{\phi}^2<0$ 
(corresponding to inverted hybrid inflation)
then we shall suppose that 
during inflation $\phi < \phi_c^-$, with the inflaton rolling away from the
origin, eventually reaching $\phi_c^-$ and ending inflation
with the same global minimum as before. Note that the global minimum
VEV $<\phi>$ is sandwiched in between $\phi_c^-$ and $\phi_c^+$
so either hybrid or inverted hybrid inflation is possible in this
model depending on the sign of $m_{\phi}^2$.

Since $A_k$ is a soft SUSY breaking parameter of order 1 TeV we have the
order of magnitude results:
\begin{equation}
k\phi_c^{\pm} \sim k<N> \sim k<\phi > \sim 1 \ TeV.
\end{equation}
Since the VEVs are associated with the large axion scale, we see that the
parameter $k\sim O(10^{-10})$. Similarly since $\lambda <N>$ plays the
role of the $\mu$ parameter of the MSSM we require
$\lambda $ to have a similarly small value. 
We shall discuss the origin of such a small
values of $\lambda, k$ later in the context of the string motivated model, 
but for now we simply note their smallness and continue.

The negative value of
$V(\phi,N)$ at the global minimum, 
is compensated by $V(0)$ which is assumed to take an equal and opposite
value, in accordance with the observed small cosmological constant.
Thus we assume:
\begin{equation}
V(0)=- V(<\phi >,<N>)= k^2<N>^4=4 k^2 (\phi_c^{\pm} - <\phi >)^4 .
\label{V0}
\end{equation}
During inflation we may set the field 
$N=0$ so that the potential simplifies to:
\begin{equation}
 V= V(0) + m^2_\phi \phi^2
\end{equation}
The slow roll conditions are given by:
\begin{equation}
\epsilon_N=\frac{1}{16\pi}\frac{M_P^2{m_\phi}^4\phi_N^2}{V(0)^2}\ll 1,
\label{epsilon}
\end{equation} 
\begin{equation}
|\eta_N | =\frac{M_P^2}{8\pi} \frac{|m_{\phi}^2|}{V(0)} \ll 1.
\label{eta}
\end{equation}
The subscripts ``N'' means 
that $\phi$ and $\epsilon$ have to be evaluated  N e-folds
before the end of inflation, when the largest scale of cosmological
interest crosses the horizon\footnote{The required number of e-folds
is roughly 60 for a potential barrier $V(0)^{1/4}\simeq 10^{16}$ GeV and
very efficient reheating in the post-inflationary period. 
It diminishes when the value of $V(0)$, and/or the reheating
temperature decrease. This will be the case for this model, 
where we will see that the needed value of $V(0)$ is lower than
$10^{11}$ GeV, and the reheating temperature is quite low. 
Nevertheless, because $|\eta |$ is extremely small,  the
particular value of $N$ is not relevant to this stage.},
that is, $N\simeq 60$. 
The height of the potential during inflation
is approximately constant and given by
$V(0)^{\frac{1}{4}}=k^{\frac{1}{2}}<N> \sim 10^8$ GeV.

Assuming that
$V(0)$ dominates the potential during inflation,
$\phi_N = \phi_c^{\pm} e^{\eta N} \approx \phi_c^{\pm}$, where the last 
approximation follows since in our model $|\eta | \ll 1/N$.
We need further to check that our inflationary model is able to
produce the correct level of density perturbation, responsible for the
large scale structure in the Universe, accordingly to the COBE
anisotropy measurements. 
The spectrum of the density perturbations is given by the quantity
\cite{deltah}, 
\begin{equation}
\delta_H^2= \frac{32}{75}\frac{V(0)}{M^4_P}\frac{1}{\epsilon_{N}}\,,
\end{equation}
with the COBE value, $\delta_H= 1.95 \times 10^{-5}$ \cite{cobe}. 
Writing $\phi_c^{\pm}\sim \phi_c$,
COBE gives the order of magnitude constraint:
\begin{equation}
 | k m_\phi | \simeq 8 \left(\frac{8 \pi}{75}\right)^{1/4} \delta_H^{-1/2}
  \frac{ (k \phi_c)^{5/2}}{M_P^{3/2}} \simeq 10^{-18} \ GeV \ 
  \left( \frac{k \phi_c}{1\, TeV} \right)^{5/2}\,.
\end{equation}
This, in turn, is more than enough to broadly satisfy the slow-roll
conditions. In particular, 
\begin{eqnarray}
|\eta_N | & 
\simeq &\frac{M_P^2}{8 \pi} \frac{|k m_\phi|^2}{(\sqrt{2} k \phi_c)^4}
\sim 10^{-12} \,, \\
\epsilon_N & \sim & \frac{M_p^2}{16 \pi} \frac{|k m_\phi|^4}{(\sqrt{2}k
\phi_c)^8} 
\phi_N^2 \sim 4 \pi \frac{ \phi_N^2}{M_P^2} \eta^2_N
\end{eqnarray}
The model predicts a very flat spectrum of density perturbations, as
usual in this type of hybrid model, with no appreciable deviation of
the spectral index, $n=1 + 2 \eta - 6 \epsilon$, from unity. Only
models where the curvature (of either sign) of the inflaton potential
is not very suppressed with respect to $H$ can give rise to a blue
\cite{bellido} (red \cite{driotto}) tilted spectrum.

Note that COBE requires the product $|km_\phi |$ to be extremely small.
If we take $k\sim 10^{-10}$, motivated by axion physics as discussed above,
then this implies $m_{\phi}$ in the electronvolt range. 
The requirement of such a small mass leads to several interesting
requirements on the model. We envisage that at the Planck scale
the $\phi$ mass is equal to zero. This can be
naturally accomplished within the framework of supergravity
no-scale models 
\cite{noscale}, where $some$  (not necessarily all) of the SUSY soft
masses are predicted to vanish, but with non-zero and universal
trilinear coupling parameters. However the high energy value of 
$m_{\phi}$ will be subject to radiative corrections which are
very small, being controlled by the small coupling $k$.
In our model the radiative corrections at $\phi \gg \phi_c$
arise from loops of the scalar and pseudoscalar components of the complex
$N$ field, which are split by soft SUSY breaking terms,
and by their fermionic partners. 
The result may be easily obtained to be \cite{coleman,shafi}:
\begin{equation}
\Delta V = \frac{k^4\phi_c^2 \phi^2}{2\pi^2}\ln(\frac{4k^2\phi^2}{\mu^2})
\end{equation} 
where $\mu$ is the modified minimal subtraction scale
and we have assumed $\phi \gg \phi_c$.
If we take $\mu^2=4k^2\phi^2 \sim 1TeV^2$  to remove the logarithm
then use the renormalisation group (RG) to run the $\phi$ mass from
the Planck scale (where it is zero) down to this scale then
the radiative corrections result in a correction to the mass of:
\begin{equation}
 \delta m_\phi^2 \simeq -\frac{|c|}{\pi^2} k^2 (k \phi_c)^2\,,
\end{equation} 
with $c$ a constant of order 1. The COBE constraint,
$k m_\phi \sim 10^{-18}$, together with the naturalness requirement
that the radiative correction is of order the mass itself,
will now translate into, 
\begin{equation}
 k \approx 10 ^{-10} \left( \frac{k \phi_c}{1\,TeV} \right)^{3/4} \,,
\label{kcobe}
\end{equation}
leading to a mass $m_{\phi}$ in the eV range.
Notice that RG equations result in a negative mass squared which would,
in the absence of any other correction, appear to favour the 
inverted hybrid inflation scenario.

However there is another potentially large contribution to the $\phi$ mass
coming from the vacuum energy $V(0)$ which breaks supersymmetry
and which drives inflation.
On general grounds, whenever local SUSY is
broken by non-zero F-terms in the visible or hidden sector, the
Kahler potential will generate
soft SUSY breaking parameters in the observable sector 
which leads to the so-called $\eta$-problem. All the scalar
fields in the observable sector will pick 
up masses of order of the Hubble constant \cite{copeland,eta1,eta2}, 
where $H^2\approx V_0 / M_P^2$, 
due to the presence of an exponential factor for
the Kahler potential in front of the
potential for the observable sector. 
The inflaton, unless fine-tuning or specific forms of the
Kahler potential are assumed \cite{eta3}, will also get this type of
mass, making it difficult to satisfy the slow-rolling constraint $\eta
\ll 1$. A linear superpotential for the inflaton is a special case
that can avoid the problem \cite{copeland} even for minimal Kahler
potential, but requires an input mass parameter. 
One simple way out of the problem is to consider $V(0)$
not having an F-term origin, but instead originating from some D-term
\cite{dinfla}. Most F--type breaking supergravity models
assume than the D-term contribution to the potential vanishes when the
fields in the hidden sector get their VEVs. 
This may not be
necessarily the case if these fields are charged under some $G_H$
hidden or visible gauge symmetry, such as a 
gauged (possibly anomalous) \ua\ 
symmetry. However the identification of $V(0)$ with some D-term seems
problematic, at least within the framework of string theories, since
the mass scale of the parameter $\xi$ predicted by string theories
is too large compared to that required by hybrid inflation \cite{riotto}, 
and in the present model this problem is made worse due to the
smallness of $V(0)$. 

The problem of the origin of $V(0)$ in our model is in fact the
cosmological constant problem and the same problem besets the
MSSM where the explicit potential at the global minimum does not
vanish, $V_{MSSM}(<H_1>,<H_2>)\neq 0$. In order to obtain a small
cosmological constant in the MSSM one is forced to
add by hand a vacuum energy $V(0)$ to accurately cancel it as in Eq.\ref{V0}.
Here, as in the MSSM, we do not specify the origin of $V(0)$, 
but simply add $V(0)$ by hand. 
The solution to the cosmological constant problem
almost certainly lies beyond supergravity, and probably beyond string theory.
What is crucial for the
success of our model is that during inflation all the F-terms
vanish, and this condition is in fact satisfied by all the explicit terms
in our model. All that we require of $V(0)$ is that it does not originate
from the F-term of a supergravity model. 
Such a vacuum energy which has the characteristics assumed here,
namely that it does not originate from the F-term of a
supergravity model, and has at least the possibility of solving the 
cosmological constant problem
has been proposed within the context of quantum cosmololgy \cite{juan}.

Having fixed the value of the Yukawa coupling, 
the vacuum expectation values
are then around the scale $\phi \sim N \sim 10^{13}$ GeV. 
More specifically, from Eq. (\ref{kcobe}) we have,
\begin{equation}
\phi_c \approx 10^{13} \left( 10^{10} k\right)^{1/3} \, GeV\,.
\end{equation}
The value of $\phi_c$ slightly exceeds the usually quoted
upper bound for the axion VEV, derived requiring the
cosmic density of axions not to exceed the critical density of the
universe \cite{kimreport}.
However a value $f_a\sim 10^{13}$ GeV is not strictly excluded when
several uncertainties entering the derivation of this bound are
allowed \cite{turner2}. 
Recently, it has been also argued \cite{kim} that a value of
$f_a$ bigger than 
$10^{12}$ GeV can be allowed in models where the reheating temperature
goes below 1 GeV, that is, below the temperature at which the axion
field begins to oscillate. The point is that during inflation 
the PQ symmetry is broken and the axion field is displaced at some
arbitrary angle, and it relaxes to zero only after reheating and only
below the QCD phase transition when its potential is tilted.
At this point the dangerous energy stored in the axion field is released,
but if the reheat temperature is of order 1 GeV then the resulting
axion density from the displaced axion field will be diluted by the entropy
release \cite{entropy} produced by the inflaton decay.
On the other hand the inflaton itself may decay directly into axions,
and this branching fraction must
be sufficiently small so that the
resulting number density of axions at the time of nucleosynthesis
is only a small fraction of a neutrino species, as we now discuss.

Immediately after inflation ends the 
masses of the singlet scalars which correspond to the oscillating mode
of the fields $N$ and $\phi$ just after they receive their global VEVs
will be $M_\phi \sim O(1\, TeV)$. The resulting ``inflaton'' mode
can decay into lighter particles, with a 
width proportional to the coupling $k^2/(4\pi)\sim (1 \ TeV)^2/f_a^2 $. 
As discussed in Ref. \cite{kim} in order to avoid conflict with 
nucleosynthesis in models where the reheat temperature is of order 1 GeV
one requires the branching fraction of the decaying inflaton into axions
not to exceed about 10\%. As pointed out \cite{kim} the inflaton coupling to
stops $m_{\tilde{t}}^2/f_a$ may dominate over the coupling to axions
$M_\phi^2/f_a$ if $m_{\tilde{t}}^2> M_\phi^2$,
providing that the stop mixing results in a kinematically accessible light
stop mass eigenstate $m_{\tilde{t1}}^2<M_\phi^2$. In this case the inflaton may
decay predominantly into stops.
The resulting decay rate may be estimated as:
\begin{equation}
 \Gamma_\phi \approx \frac{k^2}{4 \pi} M_\phi \sim 10^{-8} eV\,,
\end{equation}
which is quite suppressed with respect to, 
\begin{equation}
 H \simeq \frac{V_0^{1/2}}{3 M_P}\approx \frac{k \phi_c^2}{3 M_P} \sim
 1 MeV \,, 
\end{equation} 
The reheating temperature is given
by \cite{kolbturner}:
\begin{equation}
 T_{RH} \simeq 0.55 g_*^{-1/4} \sqrt{\Gamma_{\phi} M_P}\,,
\end{equation} 
where $g_*$ is the number of effective degrees of freedom at
reheating, and $\Gamma_{\phi}$ is the width of the inflaton decay. 
Conversion of the vacuum energy to thermal radiation through
the decay of the inflaton \phii\  into light particles will be
quite inefficient, because $\Gamma_\phi \ll H$. 
This gives a reheating temperature $T_{RH}\approx O(1-10)$ GeV.

Despite its low value, the reheat temperature
is  high enough to preserve the standard scenario for
nucleosynthesis, $T_{RH} > $  6 MeV, although quite far to allow electroweak
baryogenesis. Moreover, any pre-existing baryon asymmetry is likely to
be diluted during inflation. 
Nevertheless, as has been pointed out \cite{kim,driotto}, the amount of
baryon asymmetry needed might be produced 
directly by the decays of the inflaton. For this
mechanism to work we require the presence of baryon-number violating
operator in the superpotential, type $\lambda_{ijk}'' U_i^c D_j^c
D_k^c$. As discussed the inflaton can decay predominantly
into light stop squarks, and the subsequent decay of the stops into
two down-type quarks from this R-parity baryon number violating
operator will generate baryon-antibaryon asymmetry.
Other mechanisms, like Affleck-Dine type baryogenesis
\cite{affleckdine}, might also work.  

We now turn to the question of origin of the extremely small
values of the couplings $\lambda$ and $k$ which are required in this
model, and to whether this might be understood in terms of a deeper
theory such as string theory. In fact it 
has been argued \cite{casas,shafi2} 
that an approximate PQ symmetry that solves the strong
CP problem can arise from superstring models with exact 
discrete symmetries after compactification.
Let us assume a $Z_3 \times Z_5$ 
discrete symmetry, resulting from some string compactification 
and introduce singlets $M,\bar{M}$,
with the fields transforming as in Table 1.
All the fields are supposed to
originate from some 27 and $\overline{27}$
representations of $E_6$ apart from $\phi$ which is taken to be a singlet
of $E_6$. To be precise one can assume that 
$H_1,H_2,N,M$ originate from 27's while
$\bar{M}$ originates from a $\overline{27}$.
The superpotential is given by, 
\begin{equation}
W_{NR} = \lambda^{'} H_1 H_2 N \frac{M \mm}{M_P^2} 
-k'\phi N^2\frac{\bar{M}^2}{M_P^2} + c \frac{(M \bar{M})^3}{M_P^3}
+ d \frac{(N\bar{M})^5(M\bar{M})^2}{M_P^{11}}+\cdots
\end{equation}
where all the Yukawa couplings can be assumed to be of order unity,
and we have included only the leading physically relevant terms.
Note that terms such as $N^3$ are forbidden by the $E_6$ gauge symmetry
since the product of three 27's does not contain
three singlets. Here the $E_6$ gauge symmetry is assumed to be broken
at the string level by for example Wilson line breaking
\cite{wilson}, so questions such as doublet-triplet splitting
are addressed at the string level.
We assume that the $M, \bar{M}$ fields radiatively generate VEVs, 
as a result of a radiative mechanism due to their soft masses $m^2$
becoming negative, stabilised by F-terms arising from the above operators
with resulting VEVs
\begin{equation}
 \langle M \rangle = \langle \bar{M} \rangle 
\sim \left(\frac{mM_P^3}{c}\right)^{\frac{1}{4}}
\sim 10^{14} GeV 
\end{equation} 
As a result of these VEVs we recover the two terms of the
superpotential given in Eq. (\ref{phiNMSSM}), with
\begin{eqnarray}
\lambda & \sim &  \lambda^{'} \frac{M \mm}{M_P^2} \sim \lambda^{'}10^{-10} 
\,,\\
k & \sim &  k' \frac{\bar{M}^2}{M_P^2} \sim k'10^{-10} \,,
\end{eqnarray}
The \upq\ symmetry is explicitly
broken by the higher order term proportional to $d$.
As discussed \cite{casas} such higher order terms contribute an explicit
axion mass $\Delta m_a^2$
which tilts the axion potential slightly, and perturbs the
$\theta $ angle by an amount 
\begin{equation}
\Delta \theta \sim \frac{\Delta m_a^2f_a^2}{m_{\pi}^2f_{\pi}^2}.
\end{equation}
In order to preserve the PQ solution to the strong CP problem
we require $\Delta \theta < 10^{-8}$. Setting $d=1$ and including a 
trilinear coupling $A=10^3$ GeV, the above operator leads to
$\Delta \theta \sim 10^{-11}$ thereby preserving the PQ solution to the
strong CP problem in this model.
Note that the question of domain walls,
both coming from the breaking of the
discrete string symmetries, and those associated with axion domain walls,
which was discussed in detail in the second reference in \cite{casas},
does not arise in our model since they are both inflated away.

\begin{table}
\hfil
\begin{tabular}{ccccccc}
\hline
        & $H_1$ & $H_2$ & $N$ & $\phi$      & $M$      & $\overline{M}$ \\ 
$Z_3$   & $\alpha^2$ & $\alpha^2$ & 1  & $\alpha$ & $\alpha$ & $\alpha$   \\
$Z_5$ & 1 & 1 & 1 & $\beta$ & $\beta^3$ & $\beta^2$\\
\hline
\end{tabular}
\hfil
\caption{$Z_3 \times Z_5$ charges for the chiral supermultiplets.}
\end{table}

To summarise, we have seen that a simple variant of the NMSSM,
called $\phi$NMSSM, involving two singlets $N,\phi$
but the same number of parameters as in the original NMSSM,
opens up the posibility of solving the strong CP
problem and the $\mu$ problem, as well as providing also a mechanism for
hybrid or inverted hybrid inflation in the early Universe,
neatly side-stepping all domain wall problems.
The smallness of the couplings $\lambda$ and $k$ can be understood
in terms of a string motivated discrete symmetry, while the smallness of the
$\phi$ mass implies a no-scale supergravity origin for this parameter.
We do not specify the origin of the vacuum energy $V(0)$
which is necessary to drive inflation, and lead to an acceptably
small cosmological constant. However we do require that the source of the
vacuum energy not be the F-term of a supergravity model which would lead
to an unnacceptably large $\phi$ mass.
The magnitude of the VEVs
are of the correct order of magnitude for axion cosmology,
albeit on the upper edge of the allowed range. However the model
may provide its own cure since it has a low reheat
temperature of around 1 GeV, and the entropy produced by the inflaton
decay partially dilutes the cosmic axion density. Nevertheless, with $f_a$
just on the upper corner, one expects axion dark matter in this model.
In conclusion the $\phi$NMSSM has many interesting features and
solves several outstanding problems of particle
physics and cosmology. Like $N$, the inflaton $\phi$ resides in the visible
sector of the theory, and will mix with the two Higgs doublets
leading to the exciting possibility of experimentally observable effects
\cite{future}.

\begin{center}
{\bf Acknowledgements}
\end{center}
We would like to thank S. Abel,
M. Hindmarsh, D. Lyth, A. Riotto and J. Sanderson
for useful discussions.

\begin{center}
{\bf Note added}
\end{center}

Since our manuscript was resubmitted two further points have been brought
to our attention.
The first point raised by D.Lyth (hep-ph/9710347) is that 
M.Dine (hep-th/9207045) has showed that the model of Casas and Ross
is not viable due to the presence of additional soft operators which are
allowed by the discrete symmetry and which violate the PQ symmetry
at an unacceptable level. In our model the dangerous Dine operators such as
$NM^*\bar{M}\bar{N}^*$, and similar higher order operators,
are not present due to the absence of the $\bar{N}$ field,
and so our model is exempt from this criticism.
The second point emphasised by A.Riotto (private communication)
is that in a certain class
of no-scale supergravity models \cite{noscale} 
(those in which a Heisenberg symmetry is present)
the inflaton receives no mass of order the Hubble constant
thereby solving the $\eta$ problem \cite{eta3}.
Since we already invoke no-scale supergravity to explain the
masslessness of the inflaton, it is natural to appeal to this
mechanism in our model. 
Radiative corrections to the inflaton mass during inflation
would be negligible
due to the smallness of the couplings $k$ and $\lambda$.
This of course opens up
the possibility that the vacuum energy 
during inflation originates from F-terms after all.

\end{document}